# Pulse shape discrimination studies with a Broad-Energy Germanium detector for signal identification and background suppression in the GERDA double beta decay experiment

Dušan Budjáš[a*], Marik Barnabé Heider[a], Oleg Chkvorets[a,b], Nikita Khanbekov[a,c] and Stefan Schönert[a]

[a] *Max-Planck-Institut für Kernphysik,*
*Saupfercheckweg 1, 69117 Heidelberg, Germany*
*E-mail:* dusan.budjas@mpi-hd.mpg.de

[b] *now at: Department of Physics, Laurentian University,*
*Ramsey Lake Road, P3E 2C6 Sudbury, Ontario, Canada*

[c] *Institute for Theoretical and Experimental Physics,*
*Bolshaya Cheremushkinskaya 25, 117218 Moscow, Russia*

ABSTRACT: First studies of event discrimination with a Broad-Energy Germanium (BEGe) detector are presented. A novel pulse shape method, exploiting the characteristic electrical field distribution inside BEGe detectors, allows to identify efficiently single-site events and to reject multi-site events. The first are typical for neutrinoless double beta decays (0νββ) and the latter for backgrounds from gamma-ray interactions. The obtained survival probabilities of backgrounds at energies close to $Q_{\beta\beta}$($^{76}$Ge) = 2039 keV are $(0.93 \pm 0.08)$% for events from $^{60}$Co, $(21 \pm 3)$% from $^{226}$Ra and $(40 \pm 2)$% from $^{228}$Th. This background suppression is achieved with $(89 \pm 1)$% acceptance of $^{228}$Th double escape events, which are dominated by single site interactions. Approximately equal acceptance is expected for 0νββ-decay events. Collimated beam and Compton coincidence measurements demonstrate that the discrimination is largely independent of the interaction location inside the crystal and validate the pulse-shape cut in the energy range of $Q_{\beta\beta}$. The application of BEGe detectors in the GERDA and the Majorana double beta decay experiments is under study.

KEYWORDS: Gamma detectors; Particle identification methods.

---

[*] Corresponding author.

# Contents

## 1. Introduction

The search for neutrinoless double beta (0νββ) decay is the most powerful method to study the fundamental question whether the neutrino is its own anti-particle (Majorana particle). The absolute mass of Majorana neutrinos can then be probed down to the meV scale. One promising double beta decaying isotope is $^{76}$Ge. High-purity germanium detectors (HPGe) can be produced from germanium material enriched in $^{76}$Ge, which serves simultaneously as a source and as detector medium. Isotopic abundances of up to approximately 90% can be achieved with standard industrial enrichment techniques [1]. The benefits of the intrinsic radiopurity and excellent spectroscopic performance of HPGe detectors have been recognized early [2]. The most stringent experimental limits on 0νββ-decay have been obtained with germanium crystals enriched in $^{76}$Ge [3] and one group reported a claim of evidence [4]. Other experiments could neither confirm nor refute this claim so far [5]. The GERDA experiment [6], currently under construction at the Laboratori Nazionali del Gran Sasso in Italy, will operate bare germanium detectors, enriched in $^{76}$Ge, in liquid argon. The high-purity cryogenic liquid serves simultaneously as a coolant for the germanium detectors and as a shield against external radiation. The experiment pursues a staged implementation: The mentioned claim of evidence will be scrutinized in its first phase with about 15 kg·years exposure and a background index at $Q_{\beta\beta}$ = 2039 keV of <10$^{-2}$ counts/(keV·kg·year). The second phase of the experiment will explore half-lives of up to 2·10$^{26}$ years, with 100 kg·years of exposure and a background of <10$^{-3}$ counts/(keV·kg·year). Contingent on the results of the first two phases, a third phase is conceived to probe half-lives of >10$^{27}$ years. The corresponding effective neutrino mass of several 10 meV is predicted by neutrino oscillation experiments assuming an inverted mass hierarchy [7]. To explore this parameter regime, an exposure of several 1000 kg·years and backgrounds of <10$^{-4}$ counts/(keV·kg·year) are required. To reach such background levels, which are about three orders of magnitude below the best current values, novel techniques are required which exploit the decay characteristics and topology of 0νββ and background events.

The signature of a 0νββ-decay of $^{76}$Ge is an energy deposition of the two beta particles with an energy sum equal to $Q_{\beta\beta}$. Given the energy of the beta particles and the density of germanium crystals, the path length is of the order of one millimetre. Background signals to 0νββ-decays can arise from cosmic ray interactions, from cosmic ray induced isotopes and from natural or anthropogenic radioactivity. In particular, gamma radiation emitted in these processes can deposit energies at $Q_{\beta\beta}$ in the detector and can be miss-identified as 0νββ-decays. Backgrounds are reduced by locating such experiments deep underground and by hermetic high-purity shields. None the less, long-lived cosmic ray induced isotopes, such as e.g. $^{68}$Ge ($t_{1/2}$ = 270 days) or $^{60}$Co ($t_{1/2}$ = 5.272 years), which are produced during the germanium enrichment process or during the detector fabrication above ground, can limit the sensitivity for the 0νββ-search. Thirty days of exposure above ground leads to a $^{60}$Co background of 10$^{-3}$ counts/(keV·kg·year) at $Q_{\beta\beta}$. One efficient method to suppress this kind of background events is to exploit the event topology inside the germanium detector. $^{60}$Co events with energy deposition close to $Q_{\beta\beta}$ are characterized by multiple interaction sites internal to the crystal, dislocated from each other by up to several centimetres. Comparable event topologies arise also from the decays of $^{68}$Ge and from the interactions of external γ-ray backgrounds. Discriminating multi-site events (MSE) from single-site events (SSE) can therefore be a powerful tool to accept 0νββ-events and reject backgrounds.

The state-of-the-art to discriminate MSE from SSE events using semi- or true coaxial detectors is to analyze the signal pulse shape and to use segmented read-out electrodes [8-11]. Recently, a p-type detector with a modified read-out electrode has been developed [12] based on an earlier work of [13]. This 475 g custom designed detector exhibits superior pulse shape discrimination performance compared to coaxial detectors. The read-out electrode of this detector is miniaturized in order to lower the energy threshold by reducing the capacitive noise. It was optimized for coherent neutrino scattering and dark matter searches. Several detectors with this design were subsequently produced within the R&D efforts of the Majorana collaboration [14]. These detectors showed excellent spectroscopic performance, however it was reported that some of the experimental designs exhibited non-stable charge collection in long term operations.

Triggered by the work reported in [12], we realized that commercially available Broad Energy Germanium detectors (BEGe) [15] should have similar pulse shape discrimination fea-tures as the modified electrode detectors. Given the long-term experience of the manufacturer and their proven stable operation, BEGe detectors made of enriched germanium are of interest to the upcoming GERDA and Majorana 0νββ experiments.

This paper describes the first performance study of a BEGe detector with emphasis on MSE/SSE pulse shape discrimination (PSD) and presents a novel and robust pulse shape discrimination algorithm. It builds upon the preliminary results first published in [16].

## 2. The experimental setup

### 2.1 Features of the thick-window BEGe detector

The detector is a modified model BE5030 with a mass of 878 g, the largest BEGe commercially available from Canberra Semiconductor, N.V. Olen [17]. A schematic depiction of the crystal along with its dimensions is shown in Figure 1. The crystal is made of *p*-type HPGe with the Li-drifted *n+* contact (0.5 mm specified thickness) covering the whole outer surface, including most of the bottom part. The small *p+* contact is located in the middle of the bottom side. Typically the manufactured BEGe detectors have the top *n+* contact thickness reduced by machining to minimise low-energy γ-ray absorption in the inactive layer. This was omitted in case of the purchased diode, because it is of no advantage for ββ-decay experiments. To the contrary, thick dead layers provide a benefit of protection from alphas, betas and low energy gammas from potential surface contaminations of the detectors. The detector housing has a standard-thickness aluminium window as opposed to the low-absorption entrance windows normally supplied with BEGe detectors.

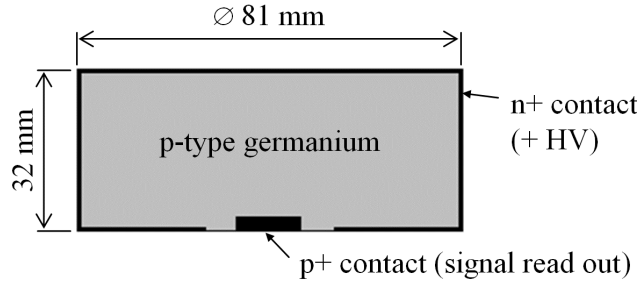

**Figure 1.** Schematic drawing of the studied 878 g BEGe-detector. Drawing adapted from [15].

### 2.2 Detector read out and data acquisition system

The front-end (FE) read out is performed with a Canberra 2002CSL RC-feedback preamplifier with a cooled FET with a specified noise level of 570 eV (at 0 pF input capacitance) and a rise time of 20 ns (at 30 pF input capacitance). The measured conversion gain is ~300 mV/MeV. The decay constant of the preamplifier signal is 47 μs. A digital data acquisition (DAQ) system is used for recording pulse shapes with 10 ns sampling rate. It consisted of a universal variable amplifier built in-house used for signal amplification without shaping and a Struck SIS 3301 14-bit 100 MHz flash ADC (FADC) with an internal trigger. The recorded trace length was 40 μs, including 9 μs of baseline. Besides recording signal traces, the FADC data-acquisition software uses on-line Gaussian filtering to shape signals and produce energy spectra. An analogue spectroscopy amplifier with an ADC system was used for dead layer and active volume determination measurements.

## 3. Detector characterisation

Full depletion and stable operation of the detector was reached at 3.6 kV bias voltage. The operating voltage was chosen at 3.8 kV. The energy resolution of the system at this voltage was 0.49 keV and 1.63 keV at the energy of 59.5 keV and 1332.5 keV, respectively (see Figure 2). Based on the reported axial gain dependence [12,13] and long-term non-stability of some modified electrode detector designs a comprehensive study to characterise the charge collection performance and stability was undertaken with the BEGe detector.

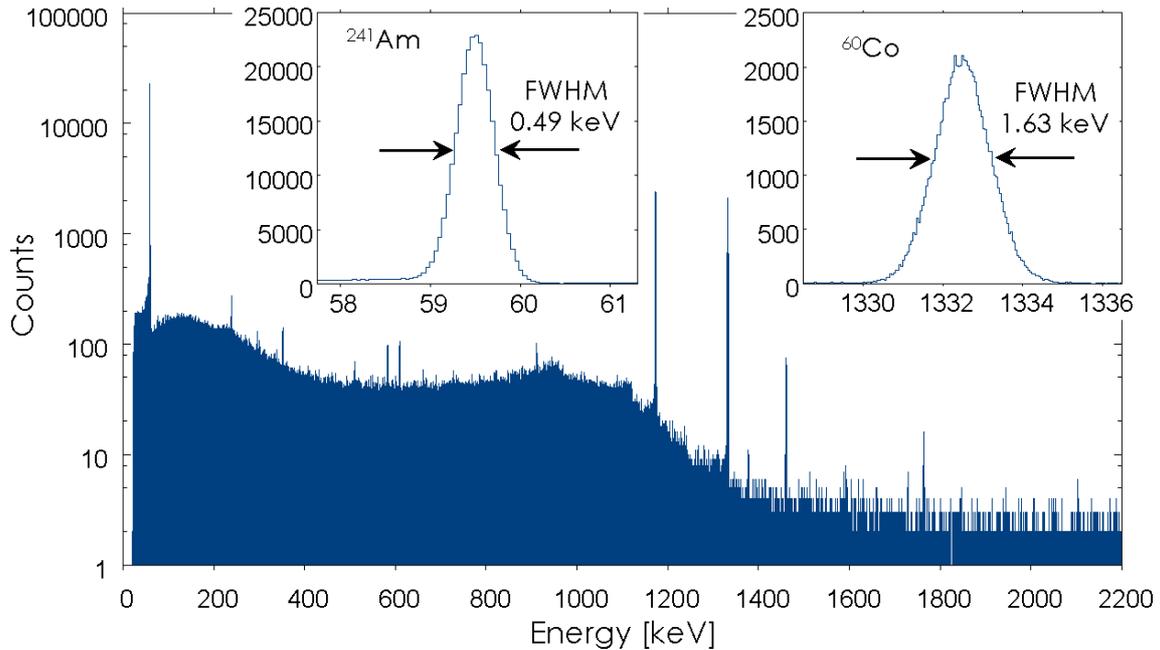

**Figure 2.** Spectrum of $^{60}$Co and $^{241}$Am obtained with the BEGe detector using 8 μs shaping with rise-time compensation. The insets show the energy resolution achieved in the 59.5 keV and 1332.5 keV peaks.

### 3.1 Experimental measurements

Measurements for the detector characterisation study were performed with the following γ-ray sources: $^{241}$Am (310 kBq) and $^{60}$Co (2.4 kBq). The $^{241}$Am source was used to perform a scan of the detector top and side surfaces with a collimated beam of low energy γ-rays. A lead collimator of 5 cm thickness with a ⌀1.5 mm hole, mounted on a precision positioning device, was used in these measurements. Three scans of the top surface along different axes and two scans at different positions along the cylindrical side of the crystal were performed. For active volume determination and long-term stability measurement, the $^{241}$Am and $^{60}$Co were measured at different distances above the detector.

### 3.2 Charge collection performance

The low energy γ-ray beam scanning was performed in order to look for charge collection losses which could occur due to inhomogeneous trapping and recombination. The maximal variation of the peak position was 0.075%, measured along the top surface. The variation was even lower in other scans and no position dependence was observed in the results. Variation in peak count rate was consistent with zero within the measurement uncertainty, which was on average ~1.5% in the top scans and ~4% in the side scans. The uncertainty was higher in the side scans due to increased γ-ray absorption by the detector holder, limiting the peak count rate. The holder shape introduced also additional systematic uncertainties to the MC simulation of the side scan. No peak tails that could be caused by charge trapping were observed in the recorded spectra [18]. The peak position and FWHM as a function of the beam position for a side scan, and the count rate variations in a top surface scan are summarised in Figure 3.

To further examine the efficiency of charge collection inside the BEGe detector, a measurement of the active volume was performed. This was done with the help of Monte Carlo simulations using MaGe [19], following a similar procedure as outlined in [20]. First, the dead layer at the top of the crystal was measured, using a ratio of count rates in detected γ-lines from an uncollimated $^{241}$Am source. The determined dead layer thickness was $(0.43 \pm 0.01)$ mm, with the dominant source of uncertainty being the thickness of the aluminium end-cap. In the second step, the active mass was evaluated using the count rate from the 1.33 MeV γ-line of a $^{60}$Co source. The thickness of surface dead layers was modified to adjust the active volume in the simulation. In this case the uncertainty was dominated by the source activity and distance. The active mass of the detector was determined to be $(833 \pm 16)$ g, 95% of the total. This would correspond to 0.46 mm of inactive material uniformly distributed on the surfaces of the detector, in good agreement with the front dead layer determination as well as with the manufacturer specification. The results of collimated beam scanning and active volume determination show that a complete charge collection is attained throughout the full detector volume.

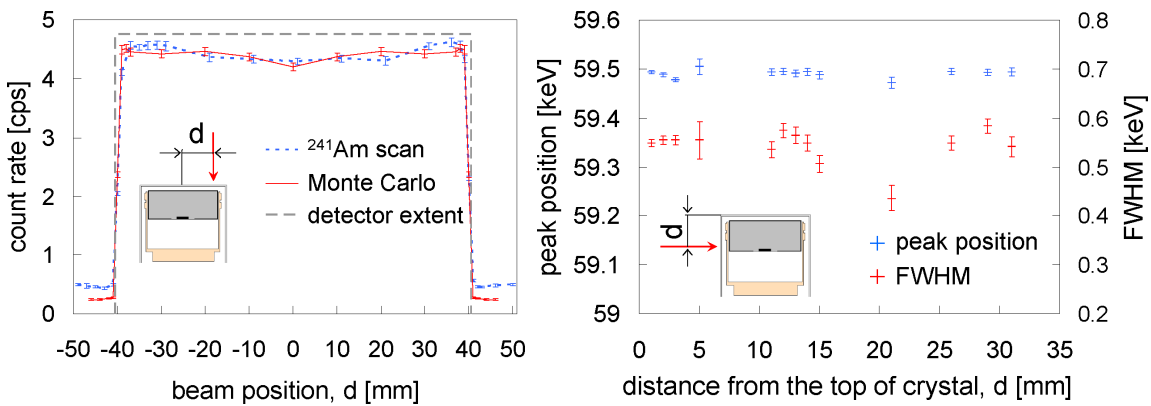

**Figure 3. Left:** Count rate variation in the 59.5 keV full energy peak along the top surface of the BEGe detector, obtained with a collimated beam. **Right:** Peak position and energy resolution variation along the side surface of the BEGe detector, obtained with a collimated 59.5 keV γ-ray beam from a $^{241}$Am source. Maximal peak position variation in side scans was 0.055%. The fluctuation of the uncertainty of the data points is caused by varying statistics due to γ-ray absorption in the detector holder.

### 3.3 Long term stability

BEGe counting characteristics were tested with two series (lasting 29 d and 22 d) of 3 hr long consecutive measurements with the $^{241}$Am and $^{60}$Co sources, and a pulse generator on the pre-amplifier test input. To correct for the gain fluctuation of the FE and DAQ system, the pulser-line position variation was subtracted from the γ-line position variation. The maximal remaining relative deviation between the pulser peak and the γ-lines was ~0.015%, corresponding to about 0.2 keV. The results of this stability test are shown in Figure 4. Correlations with liquid nitrogen (LN) refillings of the detector's dewar and with the course of the working week can be seen in the top plot. These correlations indicate that the fluctuations are caused by routine activities in the laboratory affecting the DAQ electronics and the pulser amplitude. No long-term trend can be seen in the results. The bottom half of Figure 4 shows the count rate in the 59.5 keV and 1332.5 keV peaks during the test. Overall, no evidence of fluctuation in BEGe charge collection or deterioration of the detector characteristics could be observed in the stability test. The energy resolution of the 59.5 keV, 1332.5 keV γ-lines and the pulser peak was stable at $(0.60 \pm 0.01)$ keV, $(1.62 \pm 0.02)$ keV and $(0.54 \pm 0.04)$ keV, respectively.

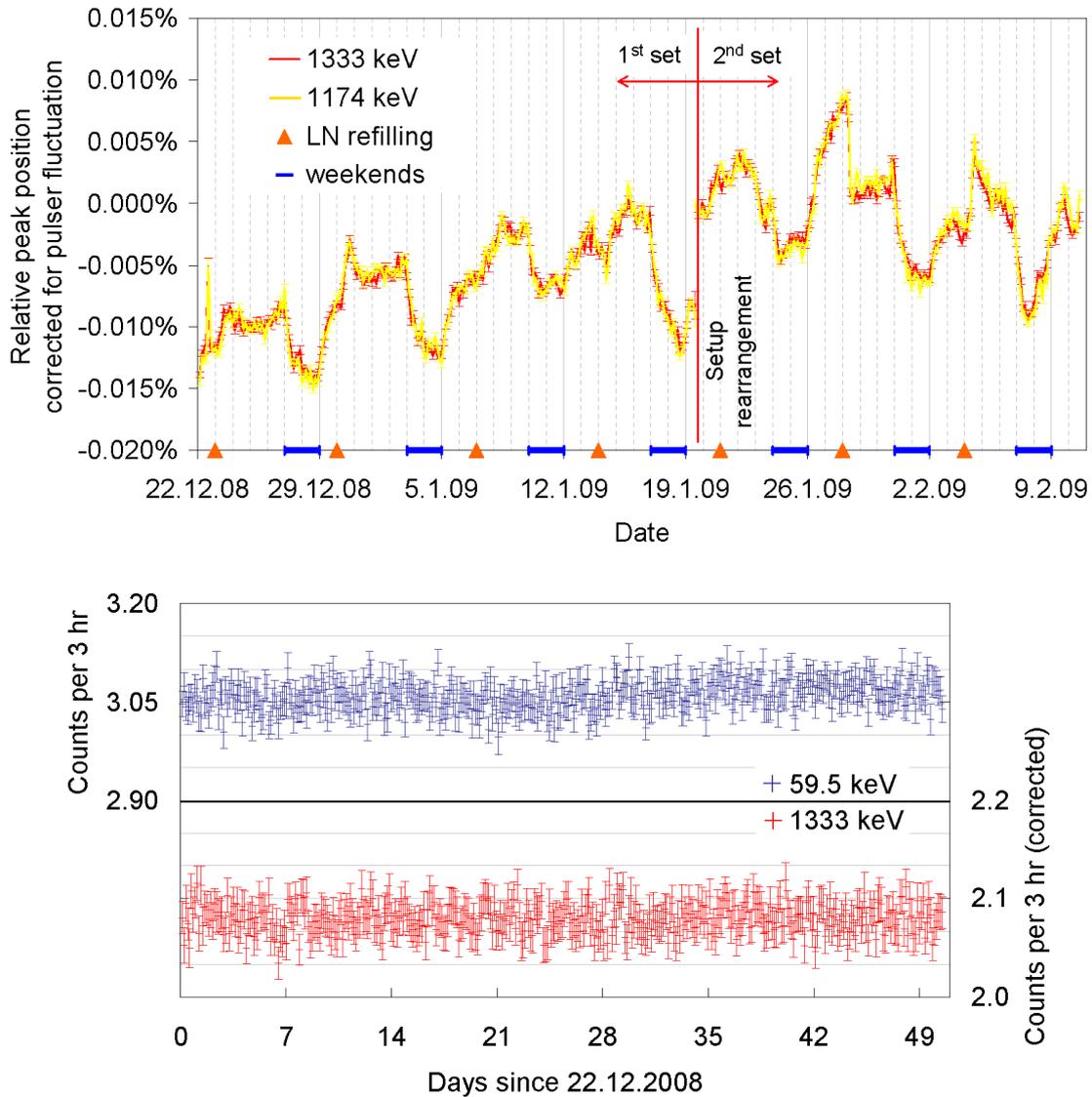

**Figure 4. Top:** Relative variation (position difference relative to the first measurement) of the 1173.2 keV and 1332.5 keV peaks, after subtracting the relative variation of the pulser peak. **Bottom:** Count rate of the 59.5 keV and 1332.5 keV peaks during the stability test. The 1332.5 keV peak count rate was corrected for the $^{60}$Co decay ($t_{1/2}$ = 5.27 y) since the beginning of the measurement.

## 4. Pulse shape discrimination method

A special feature of the BEGe detector is that a region of highly increased electric field is present near the small $p+$ electrode. Qualitatively, peaks in the output current are induced by clusters of charge carriers passing through this region [18]. For the purposes of the presented analysis, single-site events (SSE) are defined such that the charge cluster drifting through the detector has a spatial extent so small that electric field does not change significantly across its width. In case of a SSE, only one charge cluster drifts towards the $p+$ electrode and only one peak is present in the current signal (Figure 5, left). The amplitude of the induced current peak is always directly proportional to the charge contained in the cluster and in turn to the energy of the interaction that created it. In case of a multi-site event (MSE), the event energy is divided between smaller spatially separated charge clusters, which accordingly create current peaks with smaller amplitudes (Figure 5, right). These features of the BEGe current signals allow SSE/MSE discrimination based on a single parameter: the ratio $A/E$ of the maximal current signal amplitude $A$ to the event energy $E$ (proportional to the integral of the current signal). Therefore, the $A/E$ ratio of SSE should be independent of the energy and the interaction location inside the crystal volume. On the other hand, MSE are expected to have smaller $A/E$, varying with the amount of energy deposited in the largest interaction of the event.

The presented PSD technique was found to be more robust and superior to some other simple methods applied to BEGe signals, in particular counting the number of peaks in the current signal and measuring the signal rise-time. Reliable identification of small and overlapping current peaks in the presence of noise is relatively difficult to achieve, while the latter method depends on the time resolution of the FE and DAQ electronics, which was in our case 20 ns, 2.5% of a typical rise-time. On the other hand the current amplitude resolution was typically about 0.5%.

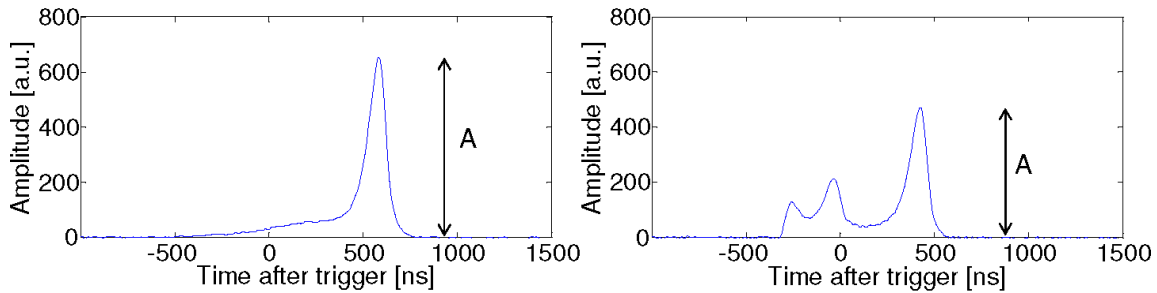

**Figure 5.** Candidate SSE (left) and MSE (right) current signals from the BEGe detector (reproduced by differentiation of recorded voltage signals), with an approximately equal energy. The maximal current amplitude $A$ is proportional to the highest energy deposition within an event.

### 4.1 Experimental measurements

A data set of $3.3 \cdot 10^7$ pulse shapes was recorded from the γ-rays of a $^{228}$Th (3.5 kBq) source, positioned at 18.5 cm from the detector end-cap. The detector current pulses were reconstructed offline by 10 ns differentiation and 50 ns smoothing of the recorded preamplifier voltage pulses. The BEGe detector was shielded by lead and copper, to minimise backgrounds from environmental radioactivity. The background spectrum is compared to the $^{228}$Th source spectrum in the top part of Figure 6.

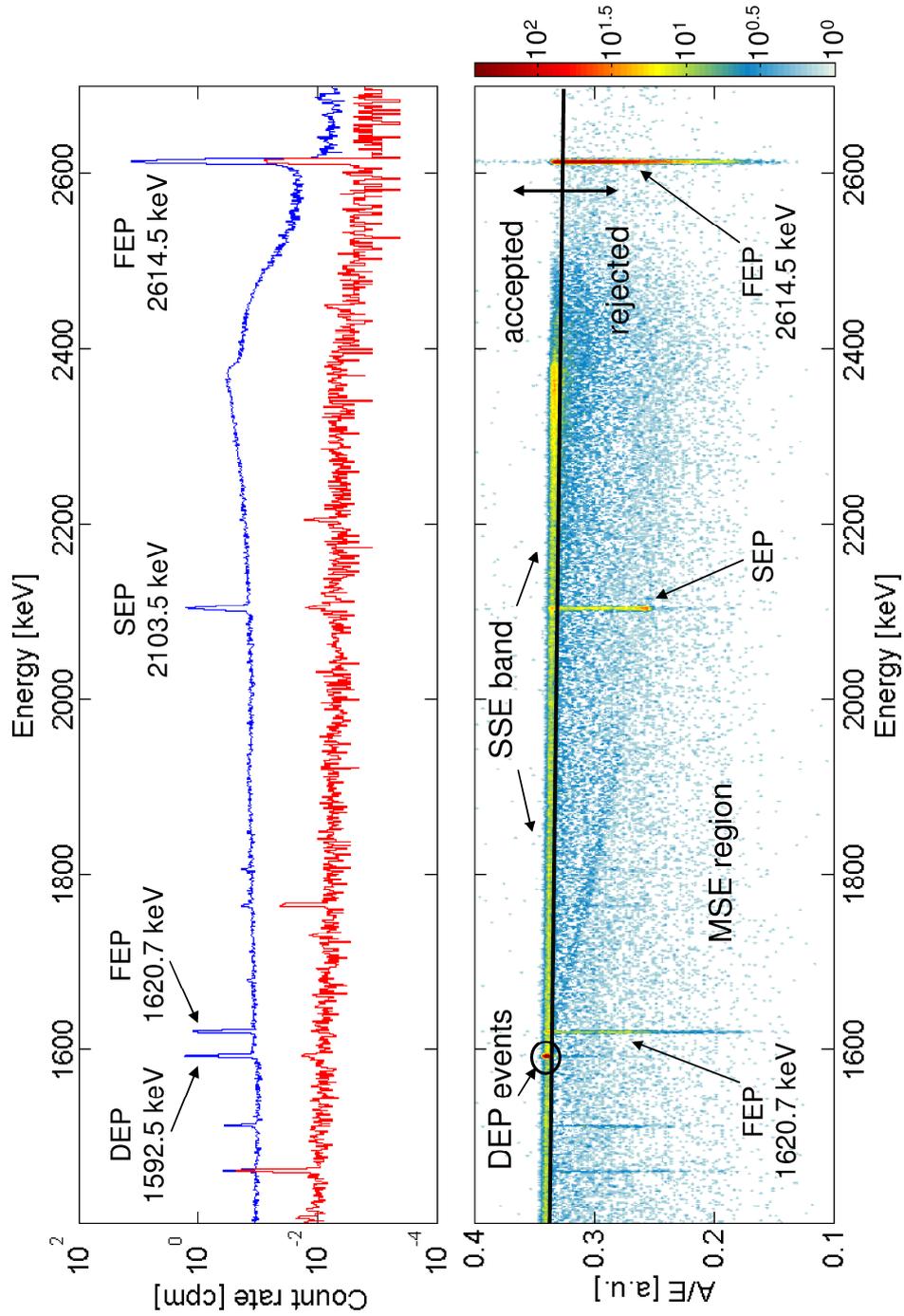

**Figure 6. Top:** $^{228}$Th spectrum (blue) recorded with the BEGe with background spectrum (red) for comparison (for the explanation of abbreviations see text). **Bottom:** Density diagram of the *A/E* parameter distribution (in arbitrary units, *a.u.*) of the same measurement, as a function of energy. The colour bar on the right indicates in logarithmic scale the number of events in a square of 1 keV × 0.0025 a.u. The thick black line shows the performed cut. The SSE band and MSE region are explained in text. Another less prominent band, visible between the DEP and SEP with a *1/E* dependence, contains pair production events of the 2614.5 keV line. The local energy deposition of the created e$^-$ and e$^+$ (constant at 1592.5 keV) is always dominant compared to the accompanying scatterings of annihilation photons, thus the current peak *A* is constant for these events, resulting in *1/E* dependence of the *A/E* values.

## 4.2 PSD calibration

The SSE/MSE discrimination cut was calibrated with the pulse shape data from the $^{228}$Th source measurements. The 1592.5 keV double-escape peak (DEP) from the 2614.5 keV emission line of $^{208}$Tl (a $^{228}$Th progeny) was used as a substitute for ββ-decay events, as it contains a dominant fraction of SSE. The absorption spectrum of $^{228}$Th features also a significant Compton continuum, which was also used in the analysis. It contains comparable fractions of SSE and MSE in the form of single-Compton scattering (SCS), respectively multiple-Compton scattering (MCS) events. The single-escape peak (SEP) of the 2614.5 keV line and the full energy absorption peaks (FEP) of $^{228}$Th emission lines (1620.7 keV and 2614.5 keV) were used as representative samples with dominant MSE content.

The distribution of *A/E* from the recorded $^{228}$Th events can be seen, together with the energy spectrum, in Figure 6. In the *A/E* density plot, the events in the Compton-continuum are separated into two groups: one is concentrated in a narrow band, approximately constant with energy; the other is spread in a broader, lower density region below the band. Following the features of BEGe pulse shapes, the high density band with constant *A/E* (approximately around *A/E* = 0.34) is assumed to contain SSE, which result from SCS events. This is highlighted in the bottom half of Figure 6 as the *SSE band*. The broader region below the band (between *A/E* = 0 and 0.33) contains MSE created by two or more times scattered γ-rays (MCS events). This area is the *MSE region* in Figure 6. This interpretation is supported by the *A/E* distribution of events in DEP, SEP and FEP. The events at the DEP energy are highly concentrated in a small spot on the SSE band. All FEP and the SEP have large tails in the MSE region.

There are also a few events above the high-density SSE band in Figure 6. These events are assumed to be caused by interactions occurring close to the read-out electrode and will be discussed in Sections 5 and 5.3.

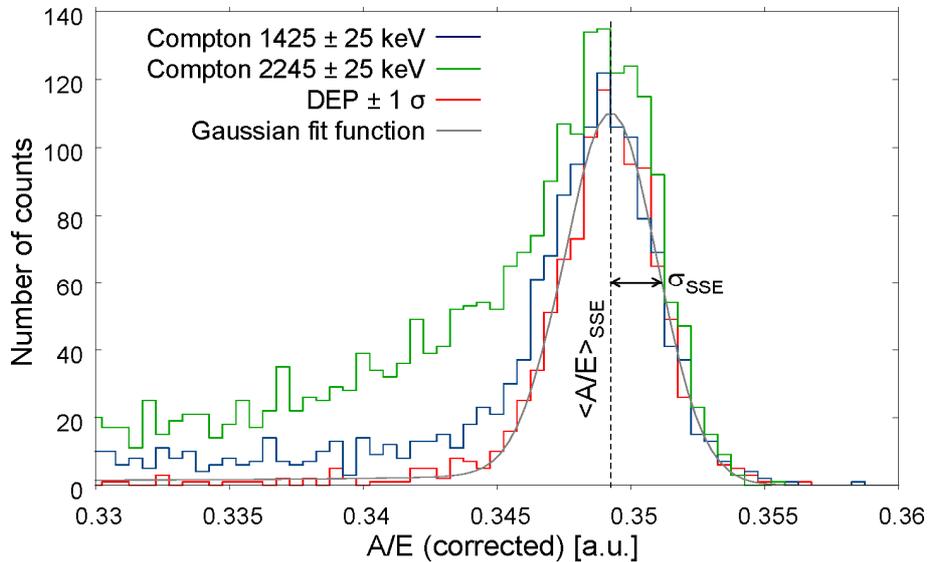

**Figure 7.** Histograms of the *A/E* distributions in cut-outs from $^{228}$Th data at different energies (DEP and two Compton regions), after a correction for the SSE band slope (see its determination in Figure 8). The histogram of the DEP events is fitted with a Gaussian function. The mean of the fit function represents the mean *A/E* value of the SSE band (*<A/E>$_{SSE}$*) and the σ its width. The Compton regions have varying MSE content, which can be approximated by an exponential tail on the left side of the Gaussian peak.

The current signal amplitude *A* essentially represents the energy of the most significant interaction in an event. Since the resolution of *A* is dominated by noise in the current signal, the *A/E* values follow a Gaussian distribution. The Gaussian shape of the SSE band cross-section can be seen in Figure 7. The MSE contribution in the *A/E* histograms shown in Figure 7 is approximated by an exponential tail. By fitting a Gaussian function with such an exponential tail to these histograms, the mean value of the SSE band ($<A/E>_{SSE}$) can be determined. The $<A/E>_{SSE}$ has a weak linear dependence on energy, which was determined from the *A/E* histograms of several Compton regions at different energies (Figure 8). The energy dependent PSD cut is selected so that the events in and above the SSE band are kept and the MSE region is rejected (see Figure 6). The acceptance of SSE can be tuned by selecting the *A/E* offset of the cut. In our analysis, the cut is performed at 2 σ below the mean of the SSE band. The Gaussian width σ was obtained from the SSE-dominated DEP data. No energy dependence of the σ width was found, therefore a constant σ value is used in the cut. Detailed description of the PSD calibration can be found in [18]. From the area of a one-sided cut of a Gaussian at 2 σ, it follows that 97.7% of SSE are accepted.

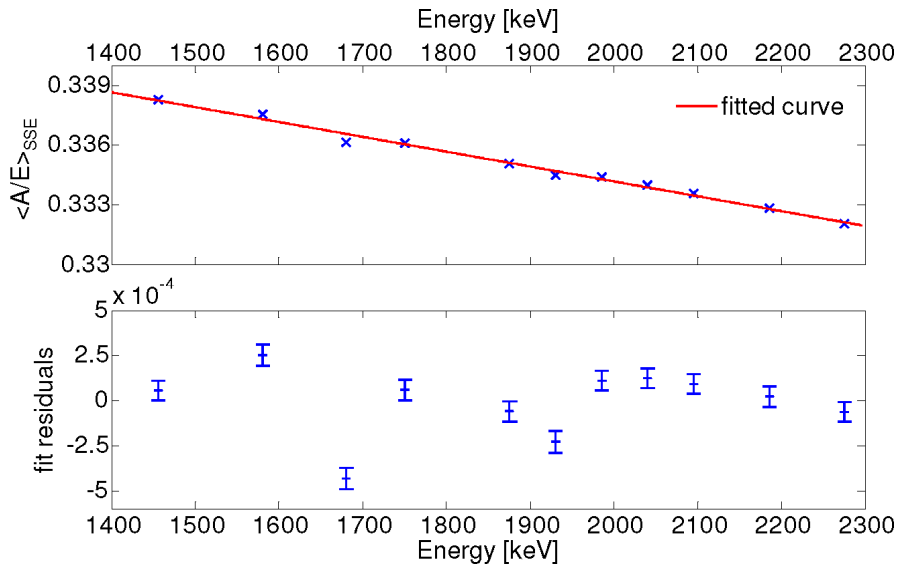

**Figure 8. Top:** Linear extrapolation of the energy dependence of the mean *A/E* values of the SSE band ($<A/E>_{SSE}$) from several Compton regions. **Bottom:** The fit residuals. The error bars include the statistical and fitting uncertainty of $<A/E>_{SSE}$.

## 5. Experimental validation of the PSD method

Coincident Compton scattering measurements allowed to test the validity of our PSD calibration based on the DEP and Compton data via a comparison with single Compton scattering (SCS) events. Similar coincidence measurements for PSD validation have been carried out in the past [8,21]. The SCS data are dominated by SSE, but have a different topology than the DEP events (single electron absorption in SCS vs. an electron and a positron in DEP) and a different spatial event distribution inside the detector. An advantage of using SCS as a sample of SSE is that they can be obtained at various energies by changing the γ-ray scattering angle. However, a disadvantage is that only a relatively low count rate can be typically achieved in coincidence measurements.

Collimated 2614.5 keV γ-ray beam measurements were performed to examine a different concern. Our PSD method so far assumed that after each interaction in the detector volume only one cluster of charge carriers passes through the region of increased electric field near the *p+* electrode. However, each interaction creates an electron cluster and a hole cluster. The electrons drift towards the *n+* electrode, therefore only holes are dominantly responsible for inducing current signals at the *p+* electrode. On the other hand, when an event occurs close to the *p+* electrode, both electrons and holes drift briefly through the region of increased electric field. In those cases their induced current adds up and the peak amplitude *A* of the induced current is higher than if only the holes would cross this region. As a consequence, the *A/E* parameter increases (events above the SSE band in Figure 6). Thus, with a fixed PSD cut value, the survival probability of MSE should increase for events with energy deposition near the electrode. In contrast, SSE should be in principle unaffected, as their *A/E* values were above the cut value already (the cut is performed only on the lower side of the *A/E* Gaussian). To measure the significance of this effect, collimator measurements were used to create a narrow beam of SSE events from DEP near the *p+* electrode of the BEGe detector.

**5.1 Coincident pulse shape measurements**

A setup with the BEGe detector in a coincidence mode with a second HPGe spectrometer (Dario, 0.83 kg active mass) was used to record SCS events (see Figure 9). The 2614.5 keV γ-rays from a 250 kBq $^{228}$Th source interact in BEGe and the scattered photons are detected by Dario. A lead and copper shield was stacked between the source and the Dario detector, to avoid a direct detection of γ-rays from the source. The source had an unobstructed view on the BEGe detector. With this arrangement, γ-ray scattering events were distributed in the whole volume of the BEGe crystal. The Dario detector was mounted on a movable table with adjustable height. By changing its position relative to the BEGe detector, the mean angle of the single-scattered γ-rays could be selected, as well as its allowed spread, which affects the width of the constrained SCS energy region. Three runs were performed, with 40º, 50º and 70º mean scattering angle, recording respectively $3.4 \cdot 10^6$, $2.9 \cdot 10^5$ and $8.9 \cdot 10^5$ events.

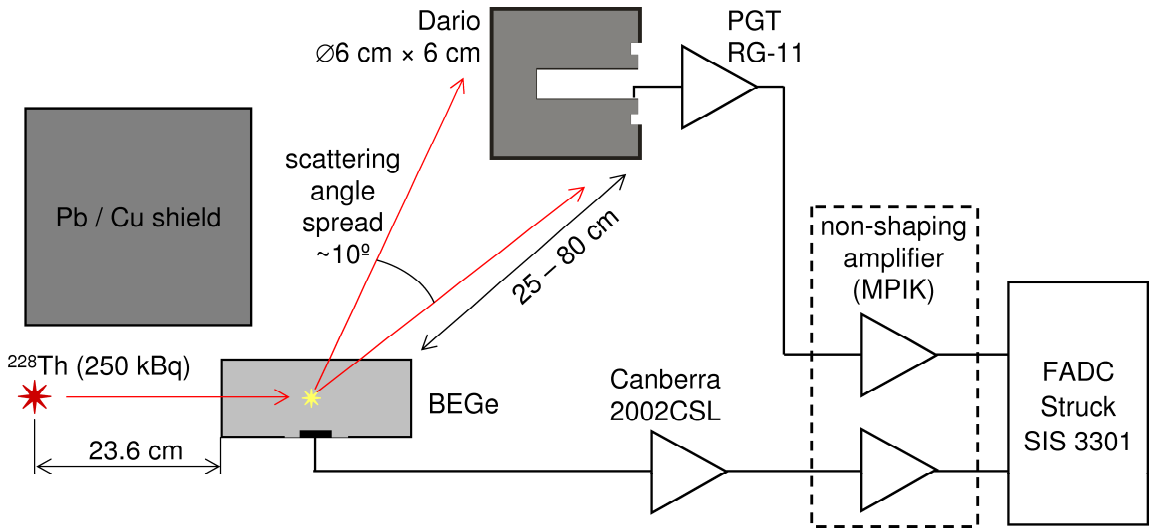

**Figure 9.** A scheme of the coincidence measurement setup, showing the electronics layout and the geometry of the measurements.

The DAQ software of the FADC triggered on events occurring in both detectors within a 4 μs wide coincidence window and a further coincidence selection was done off-line with a 1.6 μs window. The following three sets of different event classes were obtained from the recorded data for each measurement run:

1. **The full summation data set:** The scattering events of the 2614.5 keV γ-rays were selected, by requiring that a photon scattered in BEGe is fully absorbed in Dario, equalling the sum of the energy deposited in both detectors to the full 2614.5 keV (see Figure 10). This selection provided SCS events at energies corresponding to the allowed scattering angle. The width of the energy cut was on average 5 keV (~3.5 σ) on each side of the $E_{total}$ = 2.6 MeV line. Background coming from random coincidences, γ-ray cascades (e.g. a possible summation of 583 keV and 2614.5 keV γ-rays from $^{208}$Tl) and from scattering events losing partial energy outside of both detectors, is relatively small (<1%), as can be seen in Figure 10 by the different density of events in the full summation band and around it.

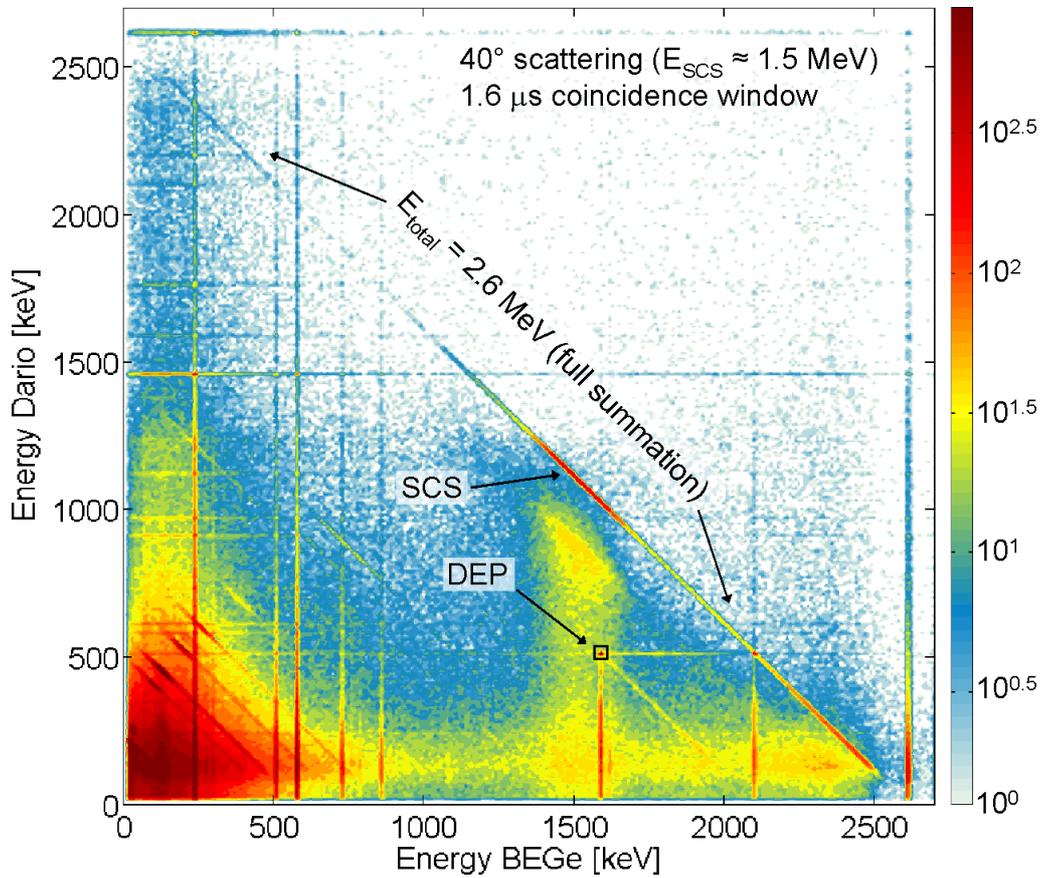

**Figure 10.** Density plot of coincident event energy in the 40° Compton scattering run. The colour bar on the right indicates in logarithmic scale the number of events in a square of 10 keV × 10 keV. The region selected for the coincident DEP data set is indicated by the black rectangle. The full summation data set was selected by taking a ~10 keV wide region around the $E_{total}$ = 2.6 MeV band. The increased density of events on the full summation band around 1.5 MeV energy in BEGe corresponds to ~40° single-Compton scattered γ-rays. The vertical and horizontal straight bands at γ-line energies occur due to random coincidences between BEGe and Dario.

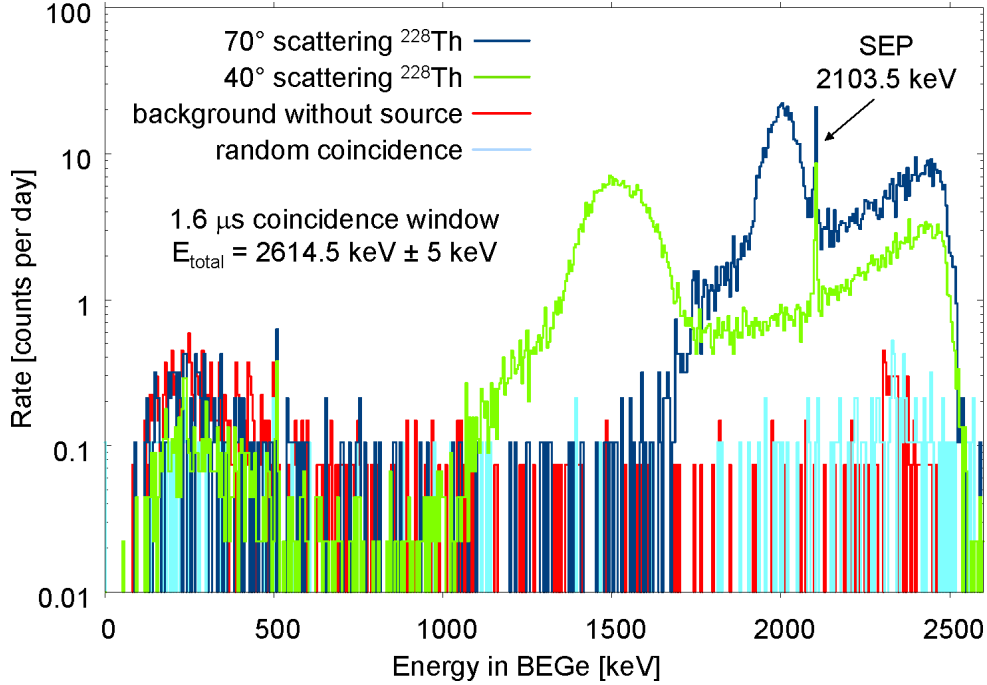

**Figure 11.** Count rate spectra of the full summation data sets from the 40º and 70º coincidence runs, compared to background contributions from random coincidences and background coincidences. The peak at 2103.5 keV is the SEP coincident with 511 keV energy depositions in Dario. Events above 2103.5 keV are dominated by events consisting of an SEP event plus an annihilation photon sharing its energy between BEGe and Dario. The random coincident data were obtained by selecting events in which the BEGe signal arrives before the 1.6 μs true coincidence window and then applying the $E_{total}$ = 2.6 MeV energy cut. The background coincidences were measured with the coincidence setup without the $^{228}$Th source, applying the same timing and energy cuts as with the SCS data.

Figure 11 shows the full summation spectra of the 40º and 70º scattering runs. The allowed spread of scattering energies was approximately (1.3 - 1.7) MeV and (1.9 - 2.1) MeV, respectively, given by (36º - 50º) and (61º - 77º) kinematically allowed ranges of scattering angles. The allowed SCS energies correspond to the broad peaks visible in the spectra, while the events at energies, which are kinematically forbidden for SCS, are created by multiple-scattered γ-rays (MCS events). These contribute MSE-dominated background to the SCS data. Contributions from the environmental background and random coincidences to the full summation data are also shown in Figure 11 and can be both neglected in the energy range of the SCS events.

**2. The coincident DEP data set:** The DEP events in coincidence with a 511 keV deposition in Dario (see Figure 10) present a SSE sample with a substantially reduced Compton background. This data set was obtained by taking a region of 3 σ on each side of the DEP/511 keV coincidence peak.

**3. The random coincidence data set:** This data set was obtained by selecting events in which the BEGe signal arrives before the 1.6 μs true coincidence window. Random coincidences mainly occur between the γ-rays from the $^{228}$Th source detected by BEGe and background events detected by Dario. This set thus represents a data sample equivalent to a non-coincident $^{228}$Th measurement and can be thus used for calibration of the energy dependent PSD cut using the Compton continuum, as is explained in Section 4.2.

## 5.2 PSD validation with coincident SCS events

The coincident DEP and random coincidence data sets were used to calibrate the PSD the same way as was done with the non-coincident measurements (see Section 4.2). The SCS data from the full summation data set were then used to test the general validity of the PSD calibration by providing SSE-rich samples at both the $^{228}$Th DEP and the $^{76}$Ge $Q_{\beta\beta}$ energies.

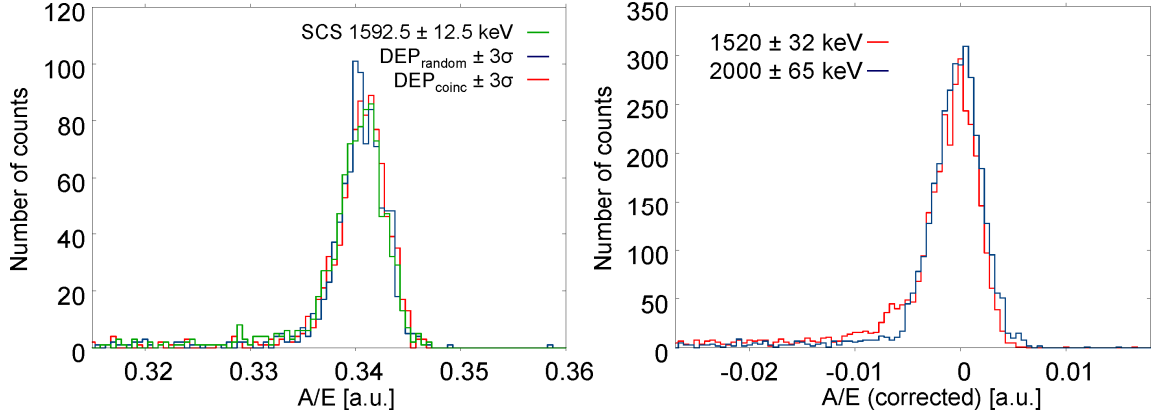

**Figure 12. Left:** *A/E* histograms of DEP (± 2.5 keV) from random and true coincident data, and SCS at 1592.5 ± 12.5 keV, all from the 40° scattering run. Each data set has an approximately equal number of events and no corrections were applied to the data. **Right:** *A/E* histograms from two SCS data sets (40° and 70° scattering), with each energy region containing an approximately equal number of events. The horizontal axis is corrected for the SSE band energy dependence and the $<A/E>_{SSE}$ value of the DEP data in each SCS set is subtracted to account for an observed *A/E* offset variation between the measurements (on the order of 1%), believed to be caused by bandwidth changes of the FE electronics. The DEP is used as a common reference, because it is present in each measurement set.

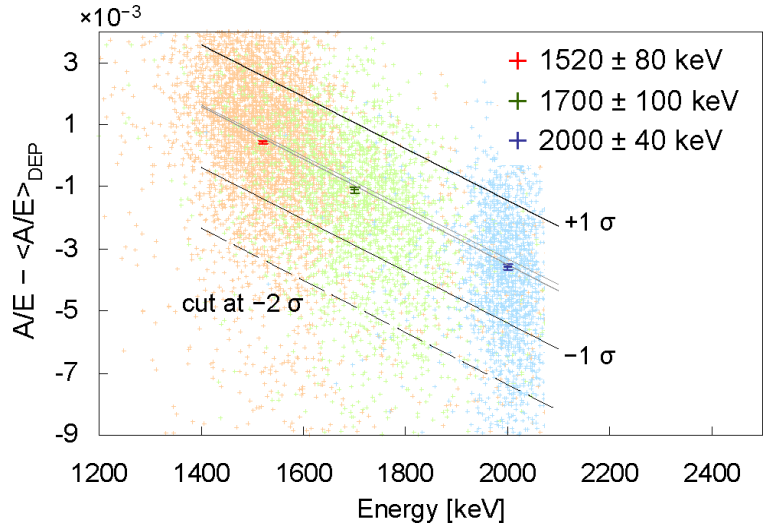

**Figure 13.** Scatter-plots of events from the 40°, 50° and 70° scattering full-summation data sets with the mean values ($<A/E>_{SSE}$) indicated. Corresponding energy intervals of the means are shown in the legend. The $<A/E>_{SSE}$ value from the DEP data ($<A/E>_{DEP}$) in each SCS set is subtracted to account for the FE electronic bandwidth variation. Superimposed is the 68% confidence interval of the SSE band mean determined from random coincidence data of the 40° run (grey lines). The 1 σ width of the SSE band and the PSD cut border line are also shown.

A comparison of *A/E* distributions from the DEP events from true and random coincident data, and from SCS events, as well as a comparison of SCS data around DEP and $Q_{\beta\beta}$ energies, is presented in Figure 12. No significant differences appear between these *A/E* distributions. Figure 13 shows a good agreement between the energy dependence of the mean *A/E* values from the SCS data and the slope of the SSE band determined from Compton regions in the random coincidence data (equivalent to a non-coincident $^{228}$Th measurement). More detailed analysis of the validation measurements can be found in [18].

## 5.3 Collimated 2614.5 keV γ-ray beam measurements

The PSD efficiency of events near the read-out electrode was studied with two pulse-shape recording measurements of a collimated γ-ray beam from the 250 kBq $^{228}$Th source. The lead collimator was 20 cm thick and featured a ⌀3 mm opening on top, increasing in steps to ⌀6 mm at the bottom. The collimated beam was aimed perpendicular to the detector end-cap, thus the events were concentrated in an approximately ⌀7 mm cylindrical volume through the thickness of the detector. One measurement was done with the collimator aimed at the centre of the crystal (6·10$^6$ events) and one aimed at ~8 mm from the edge of the crystal (8·10$^6$ events).

Figure 14 compares the *A/E* histograms of events at the DEP energy ± 1σ from the two collimated measurements and an uncollimated measurement. The histogram of the beam aimed at the detector centre has clearly more events with *A/E* above the mean value of the SSE band than the other two measurements. In the uncollimated measurement, a few events occur in this region, while with the beam near the edge, all events are concentrated on the SSE band and below. No other differences between the distributions are observed. This is in accordance with the expectation that the current peak amplitudes are amplified for interactions occurring close to the *p+* electrode, increasing the *A/E* values of such events. The consequence of this effect on the PSD results is quantified in Section 6.2.

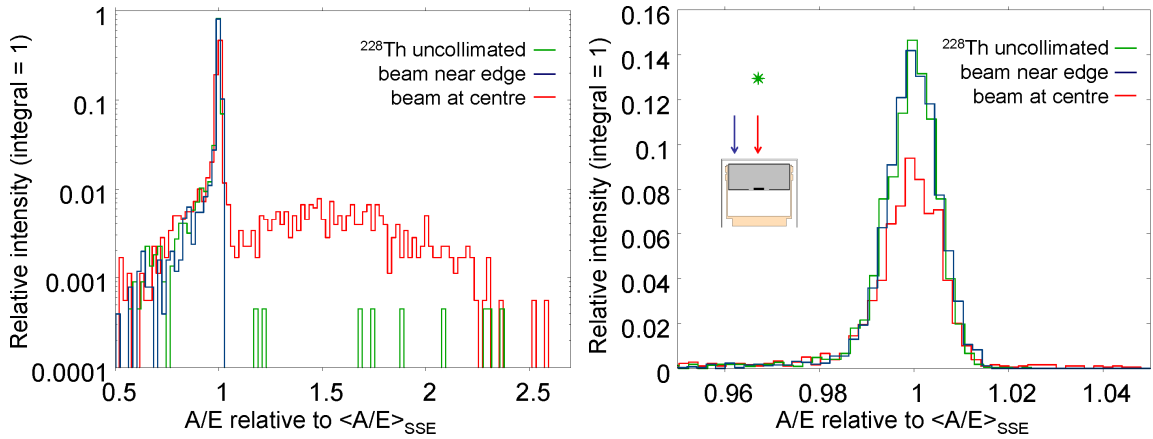

**Figure 14. Left:** Comparison of *A/E* histograms of collimated and uncollimated $^{228}$Th DEP events. The x-axis values are relative to the mean value of the SSE band (*<A/E>$_{SSE}$*), the y-axis shows the number of counts in a bin relative to the total number of counts in the whole region. **Right:** Zoom in linear scale on the SSE band mean, with a finer binning. The "beam at centre" peak is smaller, because a part of the events is shifted to the right side. The inset shows the beam locations (arrows) and the position of the uncollimated source (star), with the colours corresponding to the histograms.

# 6. 0νββ-decay background discrimination efficiency

## 6.1 Experimental measurements

Apart from the $^{228}$Th data used already for PSD calibration, pulse shapes were recorded from the γ-rays of a $^{226}$Ra (37 kBq) source measured at 70 cm above the detector end-cap and a $^{60}$Co (2.4 kBq) source at 0 cm. These data sets contained $6 \cdot 10^6$ and $3 \cdot 10^6$ pulse shapes, respectively. The $^{60}$Co data, due to the close proximity of the source contained ~20% of piled-up events, which were removed by a rejection technique based on measuring baseline variation. The BEGe detector was covered with lead and copper shielding, except in the $^{226}$Ra measurement, which consequently contained significant background in the recorded data (mostly $^{228}$Th and $^{40}$K from the surrounding materials).

## 6.2 Results and discussion

Table 1 presents the PSD results. The Compton continuum background was subtracted before and after cut in calculating the acceptances of the spectral peaks (DEP, SEP and FEP), so the values represent net acceptances. The average accepted fraction of DEP events was $(89.2 \pm 0.9)\%$. The DEP acceptance was consistent between all data sets, including the coincident and collimated beam runs. This demonstrates that the acceptance is not significantly affected by spatial distribution of events and also that the PSD method is robust with respect to numerous electronic layout changes that were performed between the individual measurements (with a recalibration of the PSD cut after each change).

A Monte Carlo simulation using the Geant4 [22] based MaGe framework [19] was performed to estimate the corresponding acceptance of 0νββ events. First, 2614.5 keV γ-rays were isotropically generated in front of the detector end-cap. About 2% of the simulated DEP events interacted via Compton scattering prior to pair production and 34% featured an emission of a bremsstrahlung photon with an energy >5 keV. The experimentally determined DEP acceptance of 89.2% can be matched in the simulation by rejecting those DEP events which have a preceding Compton interaction, or which are accompanied by a bremsstrahlung photon with an energy >90 keV. The latter condition is equivalent to an effective spatial cut, since the absorption length of 90 keV photons in germanium is 2.6 mm, thus this cut discriminates MSE. Next, $^{76}$Ge 0νββ-decays were generated uniformly in the detector active volume using the Decay0 generator [23] and subsequently propagated with MaGe. Given the higher energies involved compared to DEP, 44% of the 0νββ-events emitted bremsstrahlung photons with an energy >5 keV. Applying again a 90 keV cut to reject bremsstrahlung-accompanied events leads to an $(89.4 \pm 1.4)\%$ acceptance of 0νββ-decays. The systematic uncertainty of this simplified MC estimate is included in the result.

The table also includes survival fractions of FEPs from the $^{228}$Th source (1620.7 keV and 2614.5 keV) and the SEP of the 2614.5 keV γ-line, as well as the survival fractions of Compton events in the region around $^{76}$Ge $Q_{\beta\beta}$ from some of the most important backgrounds for 0νββ-decay ($^{228}$Th, $^{226}$Ra and $^{60}$Co). The survival probability of $^{226}$Ra events at $Q_{\beta\beta}$ is significantly smaller than that of $^{228}$Th events, because the former are dominated by the contribution from MCS interactions above the Compton edge of the 2.2 MeV line of $^{226}$Ra progeny $^{214}$Bi.

The survival probabilities of FEP and SEP events in the results from the collimated beam of $^{228}$Th γ-rays aimed at the detector centre are elevated due to the amplification of current signal amplitudes for events close to the read-out electrode (see Figure 14), as analysed in Sections 5 and 5.3. The survival fraction of the MSE-dominated SEP events in these data (~26%) can be used to approximately estimate the effective volume near the read-out electrode, in which the MSE rejection power of our PSD method is reduced. With a simplifying conservative assumption that the rejection of SEP events is inefficient only in a volume close to the electrode and perfect in the rest of the detector, the effective thickness of the volume with inefficient PSD can be estimated to be <8.2 mm from the read-out electrode (~26% of the total active volume thickness). With the diameter of the electrode of about 10 mm and assuming that the thickness of this volume is equal on all sides, a conservative limit on the effective volume with lowered sensitivity to MSE rejection by our PSD method would be <3% of the total active volume. It can be presumed, that with further investigation of BEGe pulse time structure, the MSE rejection efficiency can be improved even in this region by applying some complementary discrimination method (such as, e.g. counting the peaks in the current pulse).

**Table 1** PSD cut survival probabilities from various measurements at 97.7% SSE acceptance. The results for DEP, SEP and FEP represent net peak acceptances. The uncertainties include statistical as well as systematic uncertainties, which come from cut parameter fluctuations.

| Measurement run | DEP 1592.5 keV | FEP 1620.7 keV | SEP 2103.5 keV | FEP [a] 2614.5 keV | region near $Q_{\beta\beta}$ (2039 ± 35) keV |
|---|---|---|---|---|---|
| $^{228}$Th (set 1) | 0.889 (18) | 0.104 (12) | 0.076 (6) | 0.096 (5) | 0.408 (12) |
| $^{228}$Th (set 2) | 0.919 (30) | 0.086 (20) | 0.071 (10) | 0.103 (4) | 0.433 (16) |
| $^{228}$Th (set 3) | 0.877 (17) | 0.104 (10) | 0.066 (6) | 0.095 (4) | 0.402 (12) |
| $^{228}$Th (40° random coincidences) | 0.875 (54) | 0.162 (39) | 0.076 (22) | 0.099 (13) | 0.392 (30) |
| $^{228}$Th (40° run coincident DEP) | 0.871 (56) | – | – | – | – |
| $^{228}$Th (collimated at detector edge) | 0.913 (17) | 0.104 (15) | 0.064 (6) | 0.123 (5) | 0.464 (12) |
| $^{228}$Th (collimated at detector centre) | 0.876 (21) | 0.227 (10) | 0.258 (7) | 0.221 (4) | 0.424 (11) |
| $^{226}$Ra | – | – | – | – | 0.206 (34) [b] |
| $^{60}$Co | – | – | – | – | 0.0093 (8) |

[a] The validity of the PSD cut was tested only in the energy range between ~1.4 and ~2.4 MeV. The listed acceptance values of the 2.6 MeV peak may have a systematic error, as it lies outside this range.
[b] A contribution from background $^{228}$Th (representing 40% of events) was subtracted in the computation of this value.

Example spectra before and after the PSD cut are shown in Figure 15. Figure 16 gives the survival fractions of different spectral peaks in dependence on energy, and of $^{228}$Th, $^{226}$Ra and $^{60}$Co Compton events at $Q_{\beta\beta}$ energy of $^{76}$Ge. All full energy peaks, except for the lowest energy (1120 keV) peak, feature a constant acceptance fraction, on average (10.0 ± 0.2)%. Similar behaviour was reported from PSA measurements and simulations performed in the past [10]. The SEP has a smaller acceptance than the FEPs (7% compared to 10%) due to a larger fraction of MSE in this peak. The $^{60}$Co events in $Q_{\beta\beta}$ region (created by a partial summation of the 1.17 MeV and 1.33 MeV cascade γ-rays) and in the summation peak have a very low survival probability of 0.9%, respectively 0.3%, due to their very strong multi-site signature.

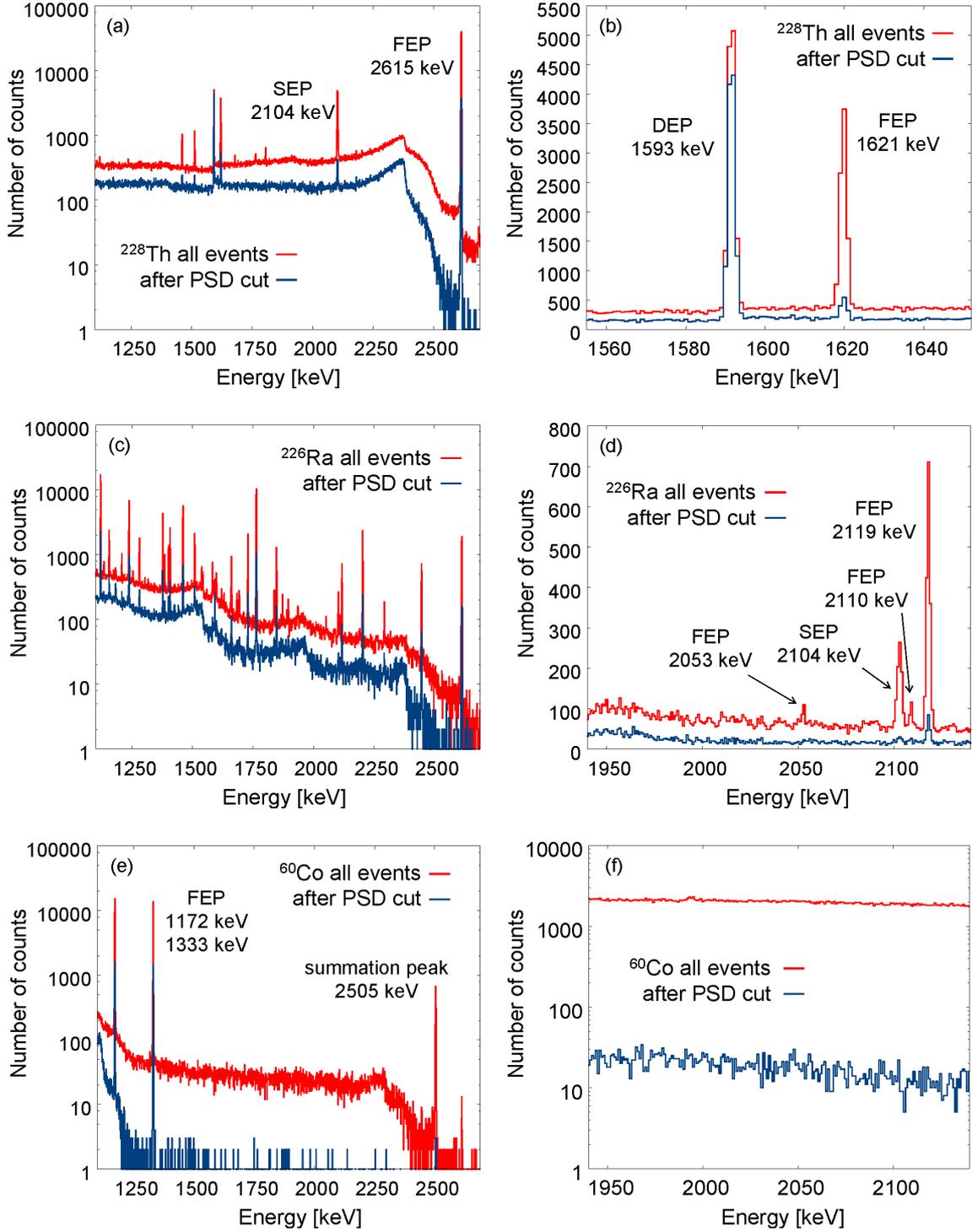

**Figure 15.** Recorded spectra from some of the most important sources of background for 0νββ-decay, before and after the PSD cut: **a)** $^{228}$Th; **b)** zoom on the DEP of the 2.6 MeV line and the 1.6 MeV FEP of $^{212}$Bi; **c)** $^{226}$Ra (with $^{228}$Th and $^{40}$K background), **d)** zoom on the Compton continuum around $^{76}$Ge $Q_{\beta\beta}$ energy. The 2.1 MeV SEP is from background $^{228}$Th; **e)** $^{60}$Co; **f)** higher statistics measurement of the Compton continuum region around $^{76}$Ge $Q_{\beta\beta}$ energy.

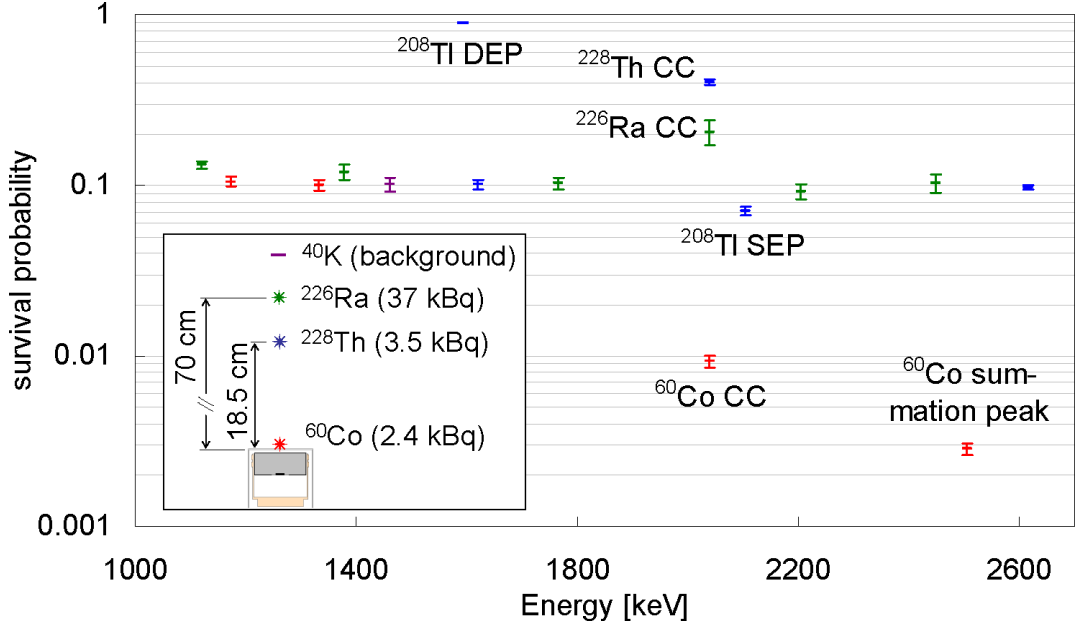

**Figure 16.** PSD cut survival probabilities of spectral peaks and events from Compton continuum (CC) around $^{76}$Ge $Q_{\beta\beta}$ energy. The data points that are not labelled are FEP. The $^{228}$Th values are weighted averages from all $^{228}$Th results except the collimated and coincidence measurements.

## 7. Conclusions and outlook

Our experimental studies carried out with an 878 g thick-window BEGe detector demonstrate its excellent performance to discriminate single site events (SSE) from multi-site events (MSE). The first are typical for double beta decays of $^{76}$Ge and the latter for γ-line and multiple-Compton event backgrounds. A novel pulse shape discrimination (PSD) method has been developed based on the ratio of the peak value of the current pulse (*A*) to the event energy (*E*). Adjusting the cut on *A/E* for 97.7% acceptance of ideal SSE, the events of the double escape peak (DEP) at 1592 keV from the 2614 keV $^{208}$Tl line are accepted in (89.2 ± 0.9)% of the cases and the events in the 1620.7 keV full energy peak have a survival probability of (10.1 ± 0.7)%. The population in the first peak is dominated by SSE and of the latter by MSE. The acceptance of $^{76}$Ge 0νββ-decay events was estimated by Monte Carlo at (89.4 ± 1.4)%.

The survival probability for γ-ray interactions from a $^{60}$Co source located on the detector cap was determined to (0.93 ± 0.08)% at $Q_{\beta\beta}$ energies, i.e. 99.1% of the events are removed. A high suppression factor is also expected for the cosmogenically produced $^{68}$Ge which has a strong MSE signature. Dedicated measurements and Monte Carlo simulations are planned for the future. It is apparent that the BEGe PSD technique can very efficiently suppress these intrinsic HPGe detector backgrounds which are one of the challenges of the GERDA and Majorana experiments.

Survival fractions at $Q_{\beta\beta}$ of (20.6 ± 3.4)% and (40.2 ± 1.6)% were determined for gamma radiation from $^{226}$Ra and $^{228}$Th respectively, the two most significant trace contaminations of materials external to the germanium crystals. The common characteristic of most external and internal backgrounds is that in order to create events at energies around $Q_{\beta\beta}$, at least one of the involved γ-rays must deposit part of its energy outside the germanium crystals. This makes the background suppression by PSD complementary to other active rejection methods, such as anti-coincidence in a detector array, or a liquid argon (LAr) scintillation veto [24]. A combined survival probability of <$10^{-4}$ for $^{60}$Co events is e.g. expected from Monte Carlo simulations when combining PSD and LAr scintillation veto.

The validation measurements with coincident Compton scattering confirmed that the DEP represents double-beta decay like events and that the PSD cut can be calibrated with the Compton continuum and the DEP from a $^{228}$Th source. It was also demonstrated that only a relatively small loss of MSE-rejection sensitivity occurs close to the read-out electrode with our PSD method. In addition, stability and charge collection investigations discovered no signs of performance degradation due to the unconventional detector configuration. The active detector volume is identical with its geometrical volume reduced by the measured dead-layer thickness. No charge collection losses were observed and the achieved energy resolution was 1.63 keV (at 1332.5 keV) and 0.49 keV (at 59.5 keV).

The novel single-parameter PSD method presented here exploits the special properties of small-electrode detectors. It leads to background rejection factors for $^{76}$Ge 0νββ-decay experiments which favourably compare with those achieved with highly segmented detectors [11]. Contaminations of signal contacts, cables and front-end electronics are one major source of external backgrounds in low-background experiments. Unsegmented detectors require fewer read-out channels than segmented detectors. Moreover, unlike segmented or point contact detectors, the unsegmented BEGe is in standard commercial production. On the other hand, BEGe detectors have the drawback that the crystals are limited in height, constraining the currently achievable mass to about 1 kg, and that a full 3D event reconstruction like in highly segmented detectors can not be realised.

As a consequence of the results of the presented investigation and the favourable characteristics of BEGe detectors, the research and development for the second phase of the GERDA experiment now includes the BEGe technology in parallel to the read-out segmentation technique.

## Acknowledgments

This work was supported in part by the Transregio Sonderforschungsbereich SFB/TR27 'Neutrinos and Beyond' by the Deutsche Forschungsgemeinschaft. Dušan Budjáš and Marik Barnabé Heider are members of the IMPRS for Astronomy & Cosmic Physics at the University of Heidelberg. We thank Dr. Jan Verplancke, Canberra Semiconductor N.V. Olen, who pointed out to us that BEGe detectors should have similar field configurations and thus pulse shape properties as those described in ref. [12]. We are also thankful to Thomas Kihm for the DAQ software and assistance with optimisation of the electronic system layout.

## Abbreviations

| | | | |
|---|---|---|---|
| 0νββ | neutrinoless double beta (decay) | HPGe | high-purity germanium |
| DAQ | data acquisition | MCS | multiple Compton scattering |
| DEP | double-escape peak | MSE | multi-site event |
| FADC | flash analogue-digital converter | PSD | pulse-shape discrimination |
| FE | front end | SCS | single Compton scattering |
| FEP | full-energy absorption peak | SEP | single-escape peak |
| FWHM | full-width at half-maximum | SSE | single-site event |